\documentclass[%
 reprint,
 amsmath,amssymb,
 aps,
]{revtex4-2}

\usepackage{graphicx}
\usepackage{dcolumn}
\usepackage{bm}
\usepackage{xcolor}
\usepackage{url}
\usepackage{lineno}

\begin{document}

\preprint{}

\title{Numerical calculation of 3-pion Coulomb scattering using scalar QED}

\author{Dhevan Gangadharan}
 \email{dhevan.raja.gangadharan@cern.ch}
\affiliation{%
 University of Houston, Houston, Texas, United States
}%


\begin{abstract}

The Coulomb scattering between three charged pions is shown to be perturbatively calculable using scalar QED for the relative momenta accessible to high-energy experiments of particle collisions.
For the triplet relative momenta, $Q_3 \gtrsim 50$ MeV/$c$, the calculated three-pion correlation function at $O(\alpha^2)$ is very similar to the commonly-used triple-product of pair factors---Riverside approximation.
At lower $Q_3$, but above the lowest existing experimental measurement of $\sim15$ MeV/$c$, terms of $O(\alpha^3)$ or higher are needed to obtain a convincing description of three-pion Coulomb interactions.
This is in contrast to two-pion correlations, for which the known exact solution is shown to be saturated by the lower orders at comparable pair relative momenta.

\end{abstract}

\maketitle

\section{\label{sec:Intro}Introduction}

The general three-body problem is well-known for its lack of a closed-form solution.
Numerical solutions are often pursued as a consequence to provide needed descriptions of three-body systems.
In the particular case of Coulomb scattering, the long-range nature of the potential presents a long-standing difficulty.
The non-relativistic quantum three-body problem can be treated using Fadeev-Merkuriev equations \cite{FadeevMerkuriev:1993} and has progressed recently \cite{Yakovlev:2016xnp, Gradusov:2021gms}.
In this letter, another approach based on perturbation theory in QED is presented.

Perturbative techniques using Feynman diagrams are suitable when the product of
the coupling and the corresponding squared matrix element is sufficiently small, thereby allowing the perturbative series to be well approximated by its first few terms.
Owing to the smallness of the fine-structure constant ($\alpha$), QED is well
suited for a perturbative approach, except in the case of atomic bound states, where
the squared matrix element can be large: $\sim 1/\alpha$.
The binding energy of two oppositely charged pions is $\sim 1/(m a^2) = 2$ keV, where $m$ is the pion mass, $a=\frac{2}{me^2}=387.5$ fm is the pion Bohr radius, and $e$ is the pion charge \cite{Lednicky:2005tb}.
The lowest relative momentum of two charged particles that is experimentally measurable in high-energy experiments of particle collisions is $\sim 5$ MeV$/c$, which is clearly well above the threshold of a $\pi^+\pi^-$ bound-state.

\begin{figure}
    \includegraphics[width=0.49\textwidth]{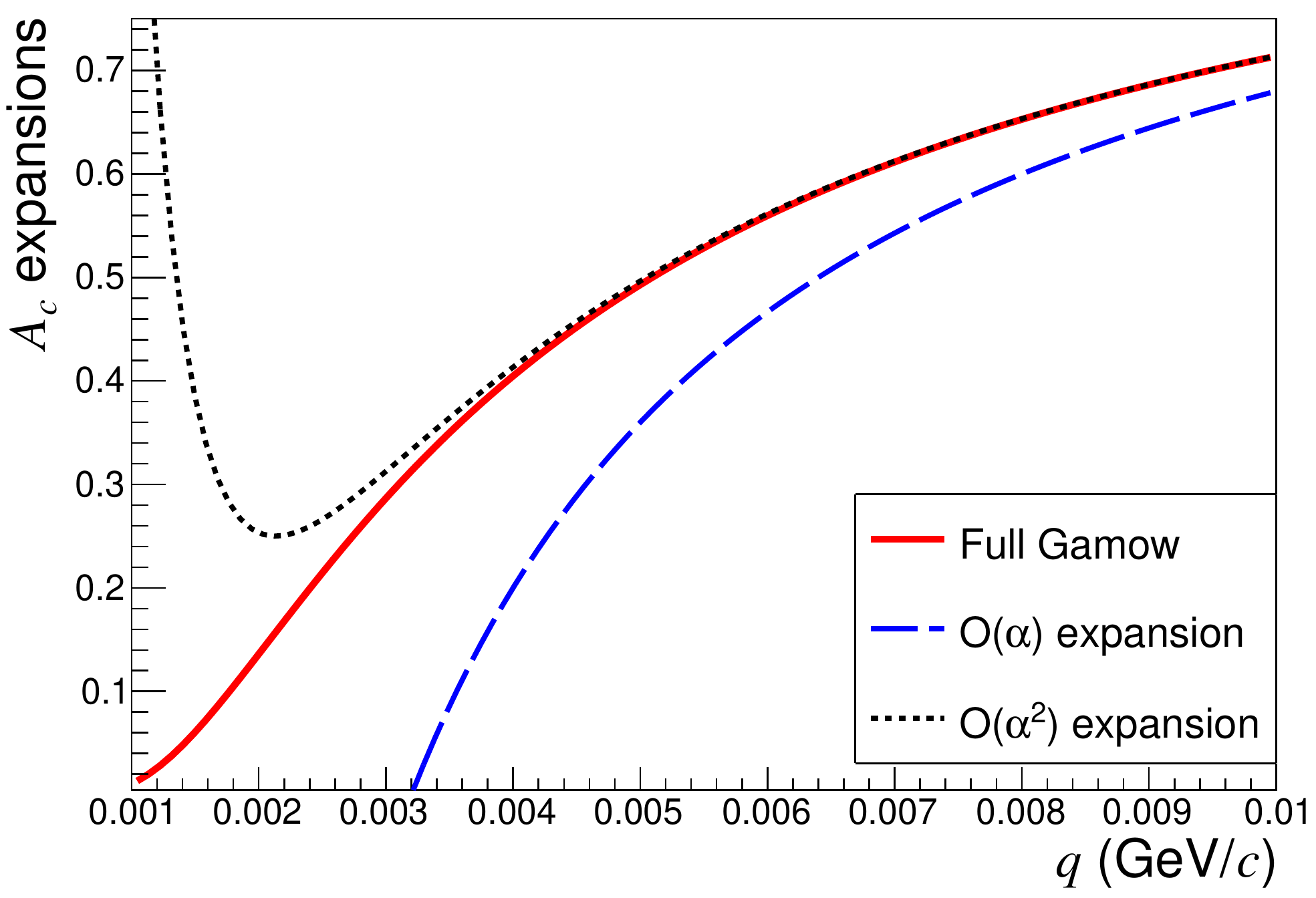}
    \caption{Gamow Coulomb factor and its expansions of $O(\alpha)$ and $O(\alpha^2)$ for identically-charged pions.}
    \label{fig:GamowExpan}
\end{figure}

In the nonrelativistic regime (low relative momentum), the problem of Coulomb scattering between two charged particles can be solved exactly \cite{Landau1981Quantum,Lednicky:2005tb}.
For the case of point-like charged particle emission, the mod square of the two-particle wavefunction is called the Gamow penetration factor and is given by
\begin{equation}
    |\Psi|^2 \equiv A_c = \frac{ 4\pi\eta }{ e^{4\pi\eta} - 1},
    \label{eq:Gamow}
\end{equation}
where $\eta = 1/(q \, a)$ and the invariant relative momentum of the pair is given by $q \equiv q_{12}=\sqrt{-(p_1^\mu - p_2^\mu)^2}$.
The same result may of course be obtained from a perturbative approach in QED and it was shown that the dominant Feynman diagram topologies are straight ladder graphs and that the entire series may be resummed analytically \cite{Baier:1969zz}.  
The mod square of the resulting amplitude was shown to coincide with the Gamow factor.
It was further pointed out that the characteristic expansion parameter in the perturbative series is given by $\alpha' \equiv \frac{\alpha \, m}{q}$, which, given the lower bound of existing measurements of $q \sim 5$ MeV$/c$, yields $\alpha' \lesssim 0.2$ for charged pions.
To illustrate the rate of convergence of the series, the Gamow factor and its expansions of $O(\alpha)$ and $O(\alpha^2)$ for identically-charged pions are shown in Fig.~\ref{fig:GamowExpan}.
One observes that for $q \gtrsim 5$ MeV$/c$, the Gamow factor is very well approximated by the $O(\alpha^2)$ expansion.
In contrast, large deviations are observed for $q < 2$ MeV$/c$, necessitating the inclusion of higher orders for smaller and smaller $q$ until eventually the problem becomes nonperturbative.

The problem of three-pion Coulomb scattering may also be solved perturbatively
using Feynman diagrams in QED provided that the triplet energy is well above
that of any bound-states.
Previous studies suggest that this condition is satisfied for the 
triplet kinematics accessible to high-energy experimental measurements \cite{Alt:1998nr}.
However, a definitive demonstration of this has not been made.
An approximation that is often used in experimental measurements is the so-called Riverside formula \cite{Liu:1986nb}, which is defined by the triple product of pair Gamow factors:
\begin{equation}
    A_c^{(3)} = A_c^{(12)} \, A_c^{(13)} \, A_c^{(23)}.
    \label{eq:Riverside}
\end{equation}
The Riverside approximation is expected to be justified in the asymptotic regime where the three particles are greatly separated in position or momentum space.

In this letter, a numerical calculation of two- and three-pion Coulomb scattering is made
using dimensionally-regulated scalar QED.  
The Feynman integral evaluation leverages recent computational
advancements in the sector decomposition method using GPUs
\cite{Binoth:2000ps,Borowka:2017idc,Borowka:2018goh}.

\section{\label{sec:ExpMeas}Observed suppression of three-pion Bose-Einstein correlations}

Correlations between identical Bosons at low relative momentum are significantly modified by quantum statistics (QS)---also known as Bose-Einstein correlations. 
The width of the correlation functions projected against relative momentum is known to be inversely related to the spatio-temporal size and shape of the particle emission region at the last stage of rescattering in particle collisions---kinetic freeze out \cite{Kopylov:1975rp,Gyulassy:1979yi,Andreev:1992pu,Lednicky:2005tb}.
A prerequisite for Bose-Einstein correlations is that some finite fraction of Boson emission occurs independently---incoherent emission. 

\begin{figure}
    \includegraphics[width=0.49\textwidth]{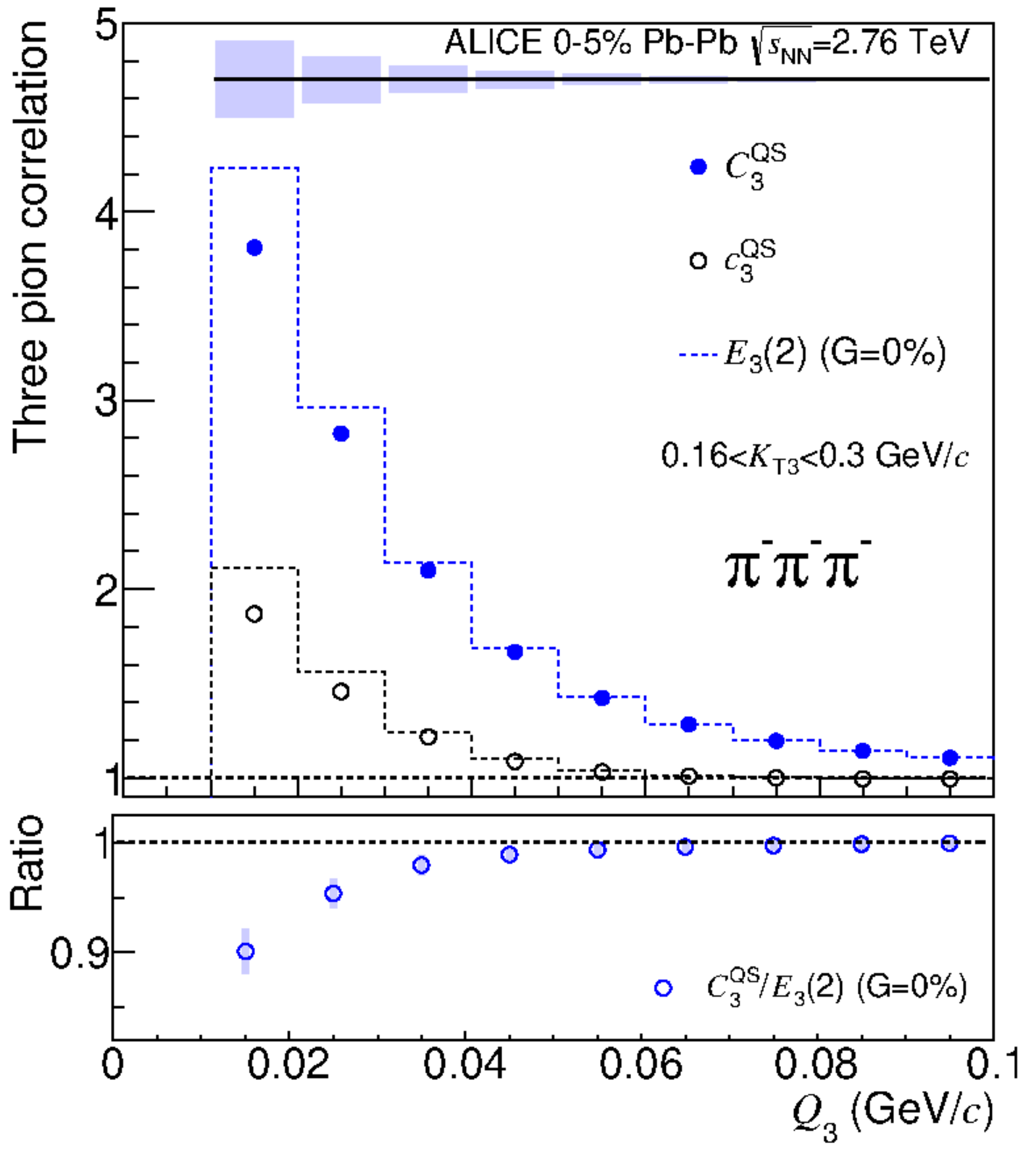}
    \caption{Three-pion QS correlations versus the triplet relative momentum $Q_3$ extracted by the ALICE collaboration \cite{ALICE:2015ryj}.  All three pions are identically charged.  Measurements are shown by the circular points and the expectations from two-pion measurements are shown as dashed lines.  Hollow circles in the top panel represent the cumulant correlation function.  A suppression of QS correlations is clearly observed for low $Q_3$.}
    \label{fig:ALICEmeas}
\end{figure}

Measurements of charged three-pion correlations at low relative momentum have been made by the ALICE collaboration \cite{ALICE:2015ryj,ALICE:2013uhj}.
The correlation function is defined as the ratio of the three-particle momentum spectra to the triple product of single-particle momentum spectra
\begin{equation}
    C_3 = \frac{N_3(p_1, p_2, p_3)}{N_1(p_1) \, N_1(p_2) \, N_1(p_3)},
    \label{eq:C3}
\end{equation}
where $p_i$ is the momentum of particle $i$.  The correlation function is usually projected against the 1D Lorentz invariant momentum measure formed by the sum quadrature of each pair invariant relative momentum: $Q_3 = \sqrt{ q_{12}^2 + q_{13}^2 + q_{23}^2 }$.
The experimental measurements are also influenced by other effects such as the dilution from particles originating from long-lived resonances which must be taken into account in order to extract the QS component \cite{ALICE:2013uhj}.
Figure \ref{fig:ALICEmeas} shows the ALICE measurements of three-pion QS correlations, $C_3^{\textrm{QS}}$. 
Also shown are the cumulant correlations, $c_3^{QS}$, which have two-body QS contributions explicitly subtracted and thereby isolate the genuine three-pion QS correlation.  
Both are compared to expectations shown with dashed lines that are built from two-pion measurements: $E_3(2)$.
The expectations are based on the assumption of fully incoherent emission, and in such a scenario, higher-order QS correlations do not contain any additional information at low $Q_3$ and are consequently fully determined by two-pion measurements.
A suppression of three-pion QS correlations with respect to the expectation is clearly observed.  

As the three pions are charged, Coulomb correlations must also be taken into account before extracting the QS component.
However, calculations that include genuine three-body effects are not available and consequently not taken into account in Fig.~\ref{fig:ALICEmeas}.
Instead, the \textit{generalized} Riverside approximation \cite{ALICE:2013uhj} was utilized, which differs from the Riverside equation (Eq.~\ref{eq:Riverside}) in that each pair factor now takes into account the finite space-time extent of particle emission. 
As mentioned, the Gamow factors, $A_c^{(ij)}$, correspond to a point-like source, which is not appropriate for the large source sizes observed in heavy-ion collisions, nor for those in $pp$ collisions \cite{ALICE:2014xrc}.
The general wavefunction for a nonrelativistic particle undergoing motion in
a Coulomb field at an arbitrary distance is known exactly \cite{Landau1981Quantum,Lednicky:2005tb}.
Assuming independent particle production, the mod square of the wavefunction may be averaged over a realistic source size, yielding the finite-size pair Coulomb factor: $K_{ij}$.
The generalized Riverside approximation to three-body Coulomb correlations is then given by: $K_3 = K^{(12)} \, K^{(13)} \, K^{(23)}$.
It is the goal of this letter to determine if the suppression seen in Fig.~\ref{fig:ALICEmeas} can be
explained by the genuine three-pion Coulomb effects that are neglected with the
generalized Riverside approach.

\section{\label{sec:Tech}Technique}

The Coulomb scattering amplitudes are calculated using Feynman diagrams in scalar QED
\cite{Schwartz:2014sze}.
Infrared (IR) and ultraviolet (UV) divergences are regulated using dimensional
regularization, where the dimensionality of each loop integral
is shifted by a small amount: $d = 4 - 2\epsilon$.
The method of sector decomposition \cite{Binoth:2000ps,Heinrich:2008si} is used to numerically evaluate the Feynman
integrals using the pySecDec package \cite{Borowka:2017idc,Heinrich:2021dbf}.
With sector decomposition, the loop integral in momentum space is first
transformed into a hypercube integral over Feynman parameters.  
The parameter integral is then split into several sectors, which allows for
a convenient means of factoring singularities of the form $1/\epsilon$ from finite regions of the
integral.
Decomposition of the integral into sectors is aided computationally by Normaliz
\cite{BRUNS20101098} and the integral in each sector is computed with the
Quasi Monte-Carlo approach using GPUs \cite{Borowka:2018goh}.

\begin{figure}
    \includegraphics[width=0.49\textwidth]{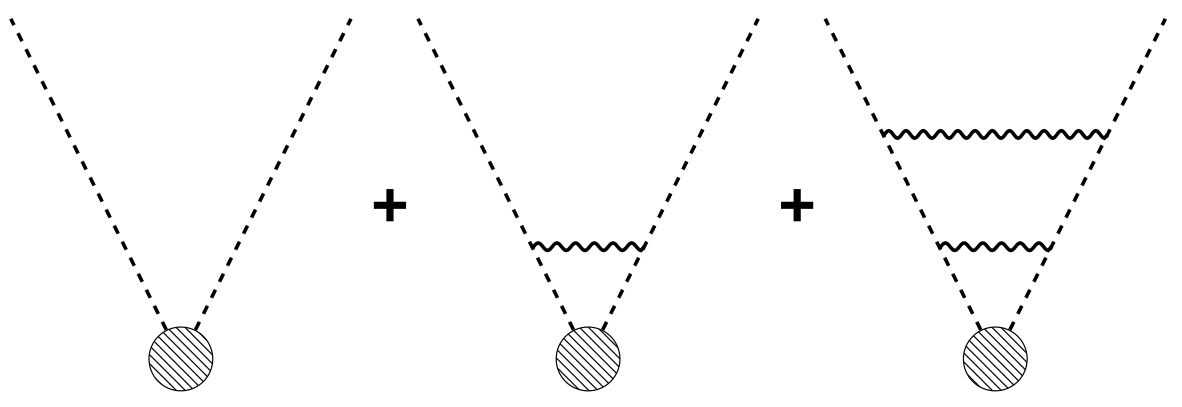}
    \caption{Feynman diagrams calculated for two-pion Coulomb correlations.  
    Diagrams made with the TikZ-Feynman tool \cite{Ellis:2016jkw}.}
    \label{fig:Diagrams2pion}
\end{figure}

There is a considerable reduction to the number of important Feynman
diagram topologies in the case of nonrelativistic final-state kinematics
\cite{Baier:1969zz}.
For diagrams containing purely virtual photons, only straight-ladder diagrams are important, which are shown in
Fig.~\ref{fig:Diagrams2pion} to order $\alpha^2$ for the two-pion case.
For the amplitudes at order $\alpha^2$ and above, non-planar diagrams where a photon passes
``underneath" another photon (non-planar) are suppressed.
Diagrams possessing four-point vertices in scalar QED ($\pi\pi\gamma\gamma$ and
$\pi\pi\pi\pi$) were calculated and found to be highly suppressed compared to
those of the same order shown in Fig.~\ref{fig:Diagrams2pion}.
Self-energy diagrams such as one where a photon begins and ends on the same pion line are
intimately related to renormalization and the associated UV divergences.
For nonrelativistic final-state particles, it was shown in Ref.~\cite{Baier:1969zz} that the loop momenta in the numerator of Feynman
integrals can be neglected, which obviates the need for renormalization.

The Feynman integrals corresponding to the diagrams in
Fig.~\ref{fig:Diagrams2pion} with $L$ loops are then given by
\begin{widetext}
\begin{equation}
  M_0 \, \left[ \alpha\mu^{2\epsilon} \left( \frac{ e^{\gamma} }{4\pi}\right)^{\epsilon} \right]^L \int\limits_{-\infty}^{\infty}  \prod_{l}^{L}  \left[ \frac{d^d k_l}{(2\pi)^d}  
  \frac{ 4 p_1 p_2 }{ [k_l^2 - i\varepsilon] \, [(p_1 + \sum_{i=1}^{l}k_i)^2 - m^2
  - i\varepsilon] \, [(p_2 - \sum_{i=1}^{l}k_i)^2 - m^2 - i\varepsilon]
  } \right].
  \label{eq:FI}
\end{equation}
\end{widetext}
The amplitude of the shaded blob is denoted by $M_0$ and represents the
hard-scattering process that created the pions in high-energy particle collisions.
For simplicity it is treated as momentum independent, which corresponds to the
point-source approximation.
It is well known that pion source sizes in hadronic collisions are, however, not
point-like, where the radii in $pp$ and $Pb-Pb$ collisions are about 1 fm and
7 fm, respectively \cite{ALICE:2014xrc}. 
Therefore, the main calculation presented here represents an upper limit to the true 
three-pion Coulomb correlation, which is to be contrasted with the Riverside approximation.
Finite-size effects can be incorporated by bringing $M_0$ inside the integral
and parametrizing its momentum
dependence in a suitable way: i.e.~ $M_0(p_1', p_2') \propto e^{-R^2
(p_1'^2 + p_2'^2)}$ \cite{Akkelin:2001nd}, where $p_i'$ is the momenta of
the pion directly emitted from the blob and $R$ is the source radius.
 
The factors raised to a power of $\epsilon$ in
Eq.~\ref{eq:FI} are a consequence of dimensional regularization of the IR
divergence that occurs when the momentum of the virtual photon goes to zero.
The parameter $\mu$ is a scale introduced to make the integral dimensionally
correct and $\gamma$ denotes Euler's constant.
Through the Kinoshita-Lee-Nauenberg theorem, the IR divergences are known to be
cancelled by the inclusion of phase-space integrals of diagrams involving real
photon emission.
Finite parts from such phase-space integrals signify the emission of soft real photons
that accompany the pions but go undetected in measurements due to finite experimental
resolution.
However, it was pointed out in Ref.~\cite{Baier:1969zz} that real-photon emission from nonrelativistic
charged particles is highly suppressed. 
Thus, the phase-space integrals are not calculated here.
The IR poles are instead subtracted in the $\overline{\textrm{MS}}$ scheme and the $\mu$
parameter is set to $m$.

\begin{figure}
    \includegraphics[width=0.49\textwidth]{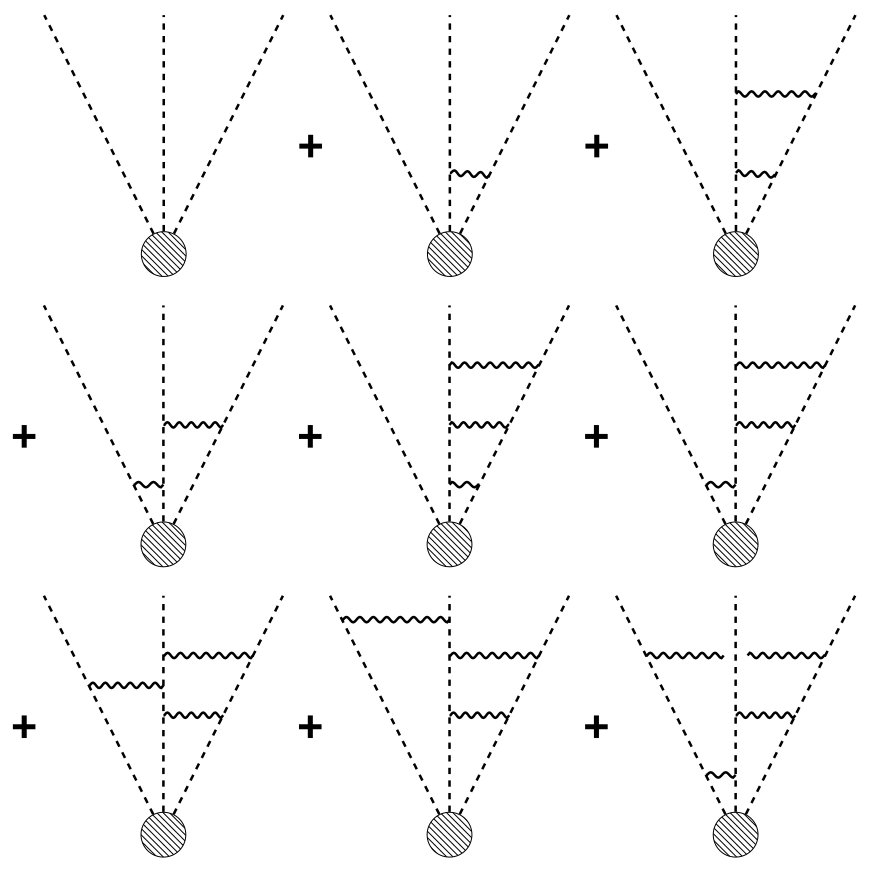}
    \caption{Feynman diagrams calculated for three-pion Coulomb correlations.  Other diagrams formed by pion permutations are included as well.  Diagrams made with the TikZ-Feynman tool \cite{Ellis:2016jkw}.}
    \label{fig:Diagrams3pion}
\end{figure}

The three-pion Feynman integrals are constructed with the same set of simplifications used for the two-pion case and the diagrams to be calculated are shown in Fig.~\ref{fig:Diagrams3pion} to $O(\alpha^3)$.
Diagrams related by pair permutations are included as well.
The compute time in pySecDec significantly increases with the number of loop integrals.
Furthermore, nonrelativistic final-state kinematics correspond to the near-threshold case, for which even larger compute times are expected. 
To reduce the complexity of the integral, integration by parts identities are utilized through the FIRE6 package \cite{Smirnov:2019qkx}, which reduces a set of integrals to a basis of simpler master integrals.
A simple set of master integrals were found only for the fourth and last diagrams in Fig.~\ref{fig:Diagrams3pion}, which resulted in a 10x reduction in compute time for those diagrams.
Further optimizations using quasi-finite bases \cite{vonManteuffel:2014qoa} could significantly reduce compute times even more, but are not pursued here.

\section{\label{sec:Res}Results}

\begin{figure}
    \includegraphics[width=0.49\textwidth]{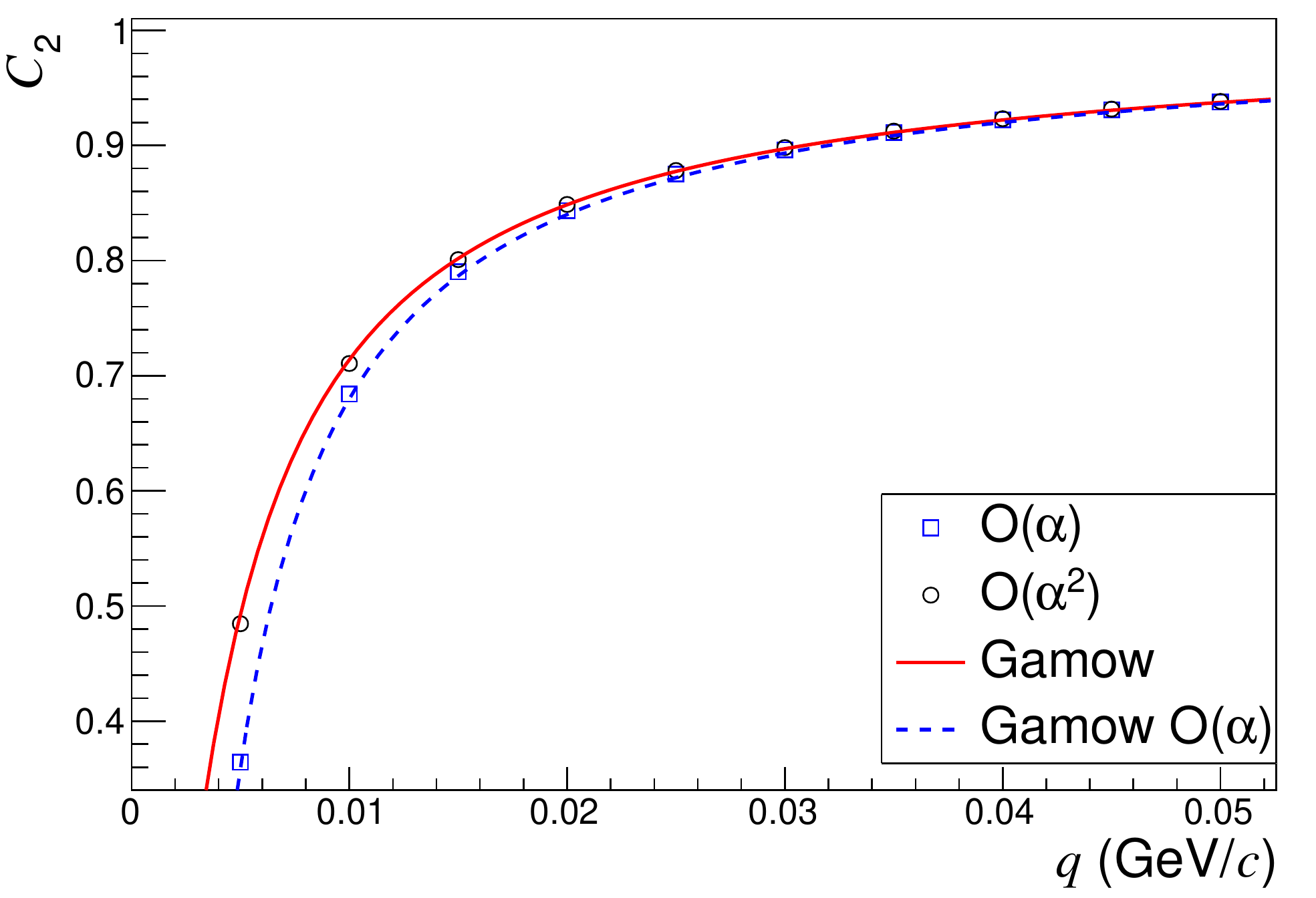}
    \caption{Two-pion Coulomb correlation function for identically charged
    pairs versus the invariant relative momentum $q$.  The calculation using
    pySecDec is shown at $O(\alpha)$ and $O(\alpha^2)$ and is compared to the
    full Gamow expectation and its $O(\alpha)$ expansion.  Statistical
    uncertainties from the numerical integration are smaller than the marker
    size.}
    \label{fig:results2pion}
\end{figure}

As the exact solution to the two-body problem of Coulomb scattering is known in the nonrelativistic limit \cite{Landau1981Quantum,Lednicky:2005tb}, the calculation of the diagrams in Fig.~\ref{fig:Diagrams2pion} serves to benchmark the techniques outlined in the previous section.
The two-pion correlation function of same-charged pairs, $C_2$, is defined as the mod square of the sum represented in Fig.~\ref{fig:Diagrams2pion} divided by the mod square of the first diagram (no scattering diagram whose mod square is $|M_0|^2$).
In the absence of correlations, $C_2=1$.
The numerical calculation at $O(\alpha)$ and $O(\alpha^2)$ with pySecDec is compared to the full Gamow expectation and its $O(\alpha)$ expansion in Fig.~\ref{fig:results2pion}.
Although an exact agreement between the Gamow and numerical calculation to $O(\alpha)$ is not expected, one should expect a high degree of similarity at the shown nonrelativistic choices of $q$.
They are consistent to better than one percent.

Three-pion correlation functions of same-charged triplets, $C_3$, are similarly defined as the ratio of the mod square of the summed amplitudes represented diagrammatically in Fig.~\ref{fig:Diagrams3pion} to the mod-square of the first diagram ($|M_0|^2$).  
The calculation is performed as a function of the triplet invariant relative momentum $Q_3$.
Unlike the pair invariant relative momentum, $q$, a specific value of $Q_3$ contains a continuum of triplet configurations, each corresponding to a different scattering amplitude.
For a point-like source emitting pions, the phase-space of a given triplet configuration is proportional to $w_3 = q_{12} \, q_{13} \, q_{23}$.
The phase-space of triplets is sampled in four steps from the most asymmetric configuration where the $q$ of one pair is at the threshold of experimental measurements (5 MeV/$c$) to the equilateral configuration where $q_{12} \sim q_{13} \sim q_{23}$.
The correlation function is then given by the $w_3$ weighted average over each triplet configuration for a given value of $Q_3$. 
As there are six pair permutations to consider for each triplet, a total of $6 \times 4 = 24$ calculations are needed for each value of $Q_3$.
At $O(\alpha^3)$, the compute time is far too large for such a spread of configurations, and at this order only the symmetric configuration ($q_{12} = q_{13} = q_{23}$) is calculated.

\begin{figure}
    \includegraphics[width=0.49\textwidth]{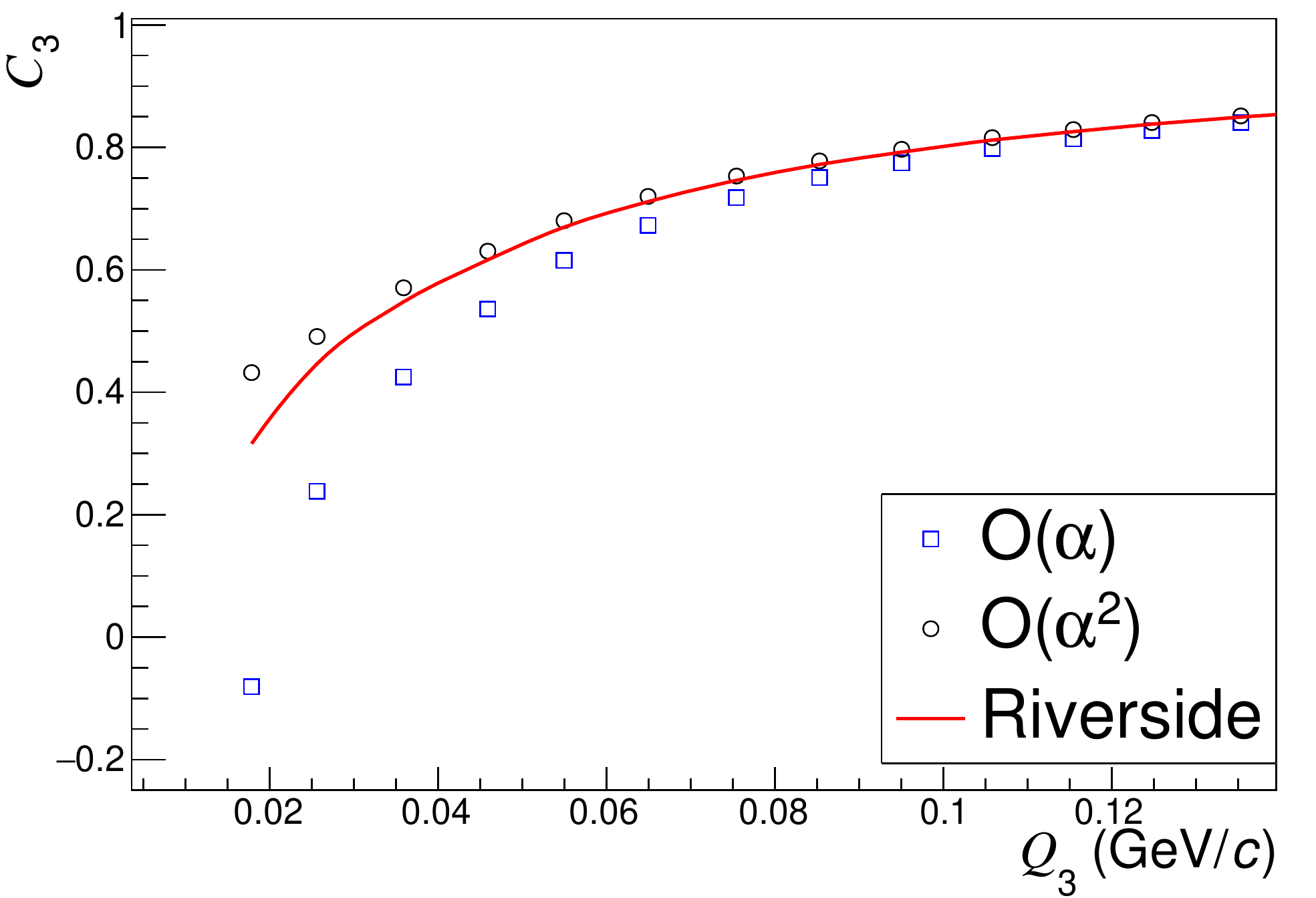}
    \caption{Three-pion Coulomb correlation function with weighted averaging for identically charged triplets versus the invariant relative momentum $Q_3$.  The calculation using pySecDec is shown at $O(\alpha)$ and $O(\alpha^2)$ and is compared to the Riverside approximation, which represents the basis of Coulomb corrections used in many experimental measurements.  Statistical
    uncertainties from the numerical integration are smaller than the marker
    size.}
    \label{fig:results3pion}
\end{figure}

\begin{figure}
    \includegraphics[width=0.49\textwidth]{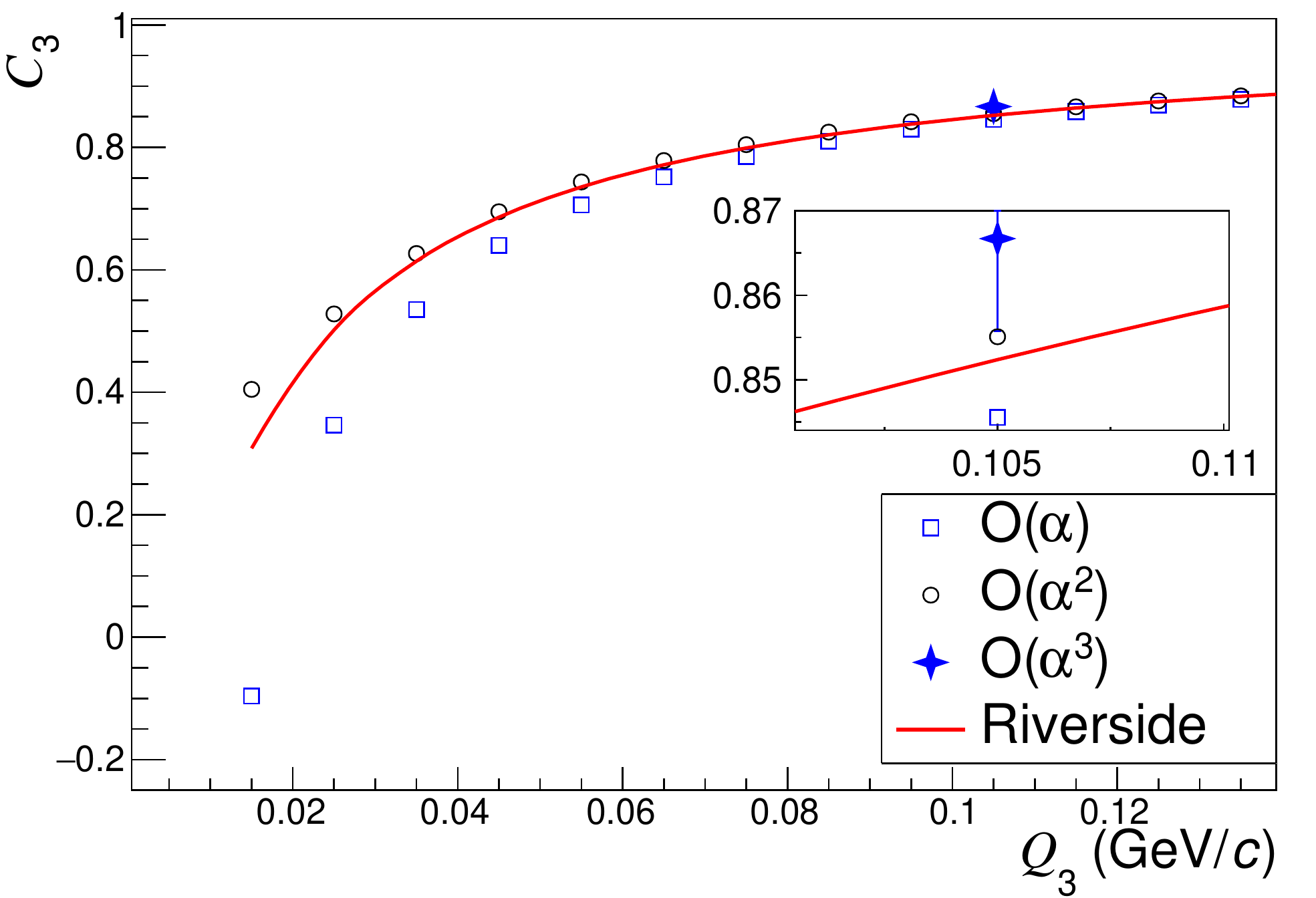}
    \caption{Three-pion Coulomb correlation function in the symmetric configuration ($q_{12} = q_{13} = q_{23}$).  The $O(\alpha^3)$ contribution is shown for only $Q_3=0.105$ GeV/$c$ due to the computational demand of the Feynman integrals near threshold.  The insert shows the calculation zoomed in at $Q_3=0.105$ GeV/$c$.  Other details are as in Fig.~\ref{fig:results3pion}.}
    \label{fig:results3pion_sym}
\end{figure}

The numerical calculation of the three-pion Coulomb correlation function using pySecDec with the weighted averaging is shown in Fig.~\ref{fig:results3pion}. 
Calculations limited to the symmetric configuration of triplets are shown in Fig.~\ref{fig:results3pion_sym}.
Shown in both Figs.~\ref{fig:results3pion} and \ref{fig:results3pion_sym} are the $O(\alpha)$ and $O(\alpha^2)$ calculations, while the $O(\alpha^3)$ results are only shown in the symmetric case for one $Q_3$ value.
Due to the increased complexity of the $O(\alpha^3)$ integrals and finite compute times, large statistical uncertainties remain at this order.
The numerical calculations are to be compared to the Riverside approximation.
Many experimental measurements use the Riverside approach as a basis for three-pion Coulomb corrections in order to extract and interpret the remaining Bose-Einstein correlations. 
In particular, the suppression of three-pion Bose-Einstein correlations observed by the ALICE experiment is dependent on this approximation.
It is therefore important to determine if the true three-pion Coulomb correlation is greater or less than that given by the Riverside approximation.
From Fig.~\ref{fig:results3pion_sym} it is seen that the sign of each order's contribution oscillates in sign, $(-1)^{\alpha}$, with decreasing magnitude.
Thus, each odd (even) order represents a new lower (upper) limit with respect to those assessed from lower orders.
Based on the relative magnitude between the $O(\alpha)$ and $O(\alpha^2)$ contributions, it can be seen that terms of $O(\alpha^3)$ or higher are needed to obtain a convincing description of three-pion Coulomb correlations at very low $Q_3$.
However, for $Q_3 \gtrsim 0.05$ GeV/$c$, the $O(\alpha^2)$ calculation appears sufficient, which is within $1\%$ of the Riverside approximation.

\section{\label{sec:Sum}Summary \& Outlook}

Two- and three-pion Coulomb interactions have been shown to be perturbatively calculable in scalar QED for the kinematics accessible to experiments involving high-energy particle collisions.
Owing to the nonrelativistic final-state kinematics of interest, a simplified set of Feynman diagrams were identified and numerically calculated using the sector decomposition method.
At the two-pion level, it is found that diagrams of $O(\alpha^2)$ are sufficient to obtain an excellent description of Coulomb scattering.
At the three-pion level, it is found that $O(\alpha^3)$ contributions or higher are needed at very low $Q_3$ but for $Q_3 \gtrsim 0.05$ GeV/$c$, $O(\alpha^2)$ appears sufficient and the often used Riverside approximation is well justified.
Since the suppression of three-pion Bose-Einstein correlations was observed for $Q_3 \lesssim 0.05$ GeV/$c$, it is not yet clear if its origin lies in extra three-body effects that are not included in the Riverside approximation. 

The calculations presented here pertain to point-like particle-emitting sources and thus represent an upper limit to the true three-body interaction in nature. 
In order to make such calculations directly applicable to experimental measurements, the finite-size effect has to be taken into account.
However, due to the large compute times of near-threshold kinematics, the additional complexity to the integration introduced by the finite size, as well as the need to sufficiently sample triplet configurations, makes an analytic approach desirable.
Since the relative momentum of pairs considered here is much smaller than the pion mass ($q_{ij}/m \ll 1$), expansion by regions or Mellin-Barnes techniques might be viable \cite{Smirnov:2012gma}.

\begin{acknowledgments}
I am very grateful to Gudrun Heinrich for taking the time to explain many details of pySecDec and perturbation theory to me.
I would also like to thank Stefan Floerchinger for his suggestions early on in this project.
I gratefully acknowledge the use of the PhysGPU cluster from the Research Computing Data Core at the University of Houston to help carry out the research presented here.
This work is supported by US DOE Nuclear Physics Grant No. DE-FG02-07ER41521.
\end{acknowledgments}

\nocite{*}

\bibliography{references}

\begin{thebibliography}{26}%
\makeatletter
\providecommand \@ifxundefined [1]{%
 \@ifx{#1\undefined}
}%
\providecommand \@ifnum [1]{%
 \ifnum #1\expandafter \@firstoftwo
 \else \expandafter \@secondoftwo
 \fi
}%
\providecommand \@ifx [1]{%
 \ifx #1\expandafter \@firstoftwo
 \else \expandafter \@secondoftwo
 \fi
}%
\providecommand \natexlab [1]{#1}%
\providecommand \enquote  [1]{``#1''}%
\providecommand \bibnamefont  [1]{#1}%
\providecommand \bibfnamefont [1]{#1}%
\providecommand \citenamefont [1]{#1}%
\providecommand \href@noop [0]{\@secondoftwo}%
\providecommand \href [0]{\begingroup \@sanitize@url \@href}%
\providecommand \@href[1]{\@@startlink{#1}\@@href}%
\providecommand \@@href[1]{\endgroup#1\@@endlink}%
\providecommand \@sanitize@url [0]{\catcode `\\12\catcode `\$12\catcode
  `\&12\catcode `\#12\catcode `\^12\catcode `\_12\catcode `\%12\relax}%
\providecommand \@@startlink[1]{}%
\providecommand \@@endlink[0]{}%
\providecommand \url  [0]{\begingroup\@sanitize@url \@url }%
\providecommand \@url [1]{\endgroup\@href {#1}{\urlprefix }}%
\providecommand \urlprefix  [0]{URL }%
\providecommand \Eprint [0]{\href }%
\providecommand \doibase [0]{https://doi.org/}%
\providecommand \selectlanguage [0]{\@gobble}%
\providecommand \bibinfo  [0]{\@secondoftwo}%
\providecommand \bibfield  [0]{\@secondoftwo}%
\providecommand \translation [1]{[#1]}%
\providecommand \BibitemOpen [0]{}%
\providecommand \bibitemStop [0]{}%
\providecommand \bibitemNoStop [0]{.\EOS\space}%
\providecommand \EOS [0]{\spacefactor3000\relax}%
\providecommand \BibitemShut  [1]{\csname bibitem#1\endcsname}%
\let\auto@bib@innerbib\@empty
\bibitem [{\citenamefont {Faddeev}\ and\ \citenamefont
  {Merkuriev}(1993)}]{FadeevMerkuriev:1993}%
  \BibitemOpen
  \bibfield  {author} {\bibinfo {author} {\bibfnamefont {L.}~\bibnamefont
  {Faddeev}}\ and\ \bibinfo {author} {\bibfnamefont {S.}~\bibnamefont
  {Merkuriev}},\ }\href
  {https://doi.org/https://doi.org/10.1007/978-94-017-2832-4} {\emph {\bibinfo
  {title} {Quantum Scattering Theory for Several Particle Systems}}}\ (\bibinfo
   {publisher} {Springer, Dordrecht},\ \bibinfo {year} {1993})\BibitemShut
  {NoStop}%
\bibitem [{\citenamefont {Yakovlev}(2016)}]{Yakovlev:2016xnp}%
  \BibitemOpen
  \bibfield  {author} {\bibinfo {author} {\bibfnamefont {S.~L.}\ \bibnamefont
  {Yakovlev}},\ }\bibfield  {title} {\bibinfo {title} {{Asymptotic behavior of
  the wave function of three particles in a continuum}},\ }\href
  {https://doi.org/10.1134/S0040577916010116} {\bibfield  {journal} {\bibinfo
  {journal} {Theor. Math. Phys.}\ }\textbf {\bibinfo {volume} {186}},\ \bibinfo
  {pages} {126} (\bibinfo {year} {2016})}\BibitemShut {NoStop}%
\bibitem [{\citenamefont {Gradusov}\ \emph {et~al.}(2021)\citenamefont
  {Gradusov}, \citenamefont {Roudnev}, \citenamefont {Yarevsky},\ and\
  \citenamefont {Yakovlev}}]{Gradusov:2021gms}%
  \BibitemOpen
  \bibfield  {author} {\bibinfo {author} {\bibfnamefont {V.~A.}\ \bibnamefont
  {Gradusov}}, \bibinfo {author} {\bibfnamefont {V.~A.}\ \bibnamefont
  {Roudnev}}, \bibinfo {author} {\bibfnamefont {E.~A.}\ \bibnamefont
  {Yarevsky}},\ and\ \bibinfo {author} {\bibfnamefont {S.~L.}\ \bibnamefont
  {Yakovlev}},\ }\bibfield  {title} {\bibinfo {title} {{Theoretical Study of
  Reactions in the ${{e}^{ - }}{{e}^{ + }}\bar {p}$ Three Body System and
  Antihydrogen Formation Cross Sections}},\ }\href
  {https://doi.org/10.1134/S0021364021130026} {\bibfield  {journal} {\bibinfo
  {journal} {JETP Lett.}\ }\textbf {\bibinfo {volume} {114}},\ \bibinfo {pages}
  {11} (\bibinfo {year} {2021})},\ \Eprint {https://arxiv.org/abs/2105.03786}
  {arXiv:2105.03786 [physics.atom-ph]} \BibitemShut {NoStop}%
\bibitem [{\citenamefont {Lednicky}(2009)}]{Lednicky:2005tb}%
  \BibitemOpen
  \bibfield  {author} {\bibinfo {author} {\bibfnamefont {R.}~\bibnamefont
  {Lednicky}},\ }\bibfield  {title} {\bibinfo {title} {{Finite-size effects on
  two-particle production in continuous and discrete spectrum}},\ }\href
  {https://doi.org/10.1134/S1063779609030034} {\bibfield  {journal} {\bibinfo
  {journal} {Phys. Part. Nucl.}\ }\textbf {\bibinfo {volume} {40}},\ \bibinfo
  {pages} {307} (\bibinfo {year} {2009})},\ \Eprint
  {https://arxiv.org/abs/nucl-th/0501065} {arXiv:nucl-th/0501065} \BibitemShut
  {NoStop}%
\bibitem [{\citenamefont {Landau}\ and\ \citenamefont
  {Lifshitz}(1981)}]{Landau1981Quantum}%
  \BibitemOpen
  \bibfield  {author} {\bibinfo {author} {\bibfnamefont {L.~D.}\ \bibnamefont
  {Landau}}\ and\ \bibinfo {author} {\bibfnamefont {L.~M.}\ \bibnamefont
  {Lifshitz}},\ }\href {http://www.worldcat.org/isbn/0750635398} {\emph
  {\bibinfo {title} {Quantum Mechanics Non-Relativistic Theory, Third Edition:
  Volume 3}}},\ \bibinfo {edition} {3rd}\ ed.\ (\bibinfo  {publisher}
  {Butterworth-Heinemann},\ \bibinfo {year} {1981})\BibitemShut {NoStop}%
\bibitem [{\citenamefont {Baier}\ and\ \citenamefont
  {Fadin}(1970)}]{Baier:1969zz}%
  \BibitemOpen
  \bibfield  {author} {\bibinfo {author} {\bibfnamefont {V.~N.}\ \bibnamefont
  {Baier}}\ and\ \bibinfo {author} {\bibfnamefont {V.~S.}\ \bibnamefont
  {Fadin}},\ }\bibfield  {title} {\bibinfo {title} {{Coulomb Interaction in the
  Final State}},\ }\href@noop {} {\bibfield  {journal} {\bibinfo  {journal}
  {JETP}\ }\textbf {\bibinfo {volume} {30}},\ \bibinfo {pages} {127} (\bibinfo
  {year} {1970})},\ \bibinfo {note}
  {\url{http://www.jetp.ras.ru/cgi-bin/e/index/e/30/1/p127?a=list}}\BibitemShut
  {NoStop}%
\bibitem [{\citenamefont {Alt}\ \emph {et~al.}(1999)\citenamefont {Alt},
  \citenamefont {Csorgo}, \citenamefont {Lorstad},\ and\ \citenamefont
  {Schmidt-Sorensen}}]{Alt:1998nr}%
  \BibitemOpen
  \bibfield  {author} {\bibinfo {author} {\bibfnamefont {E.~O.}\ \bibnamefont
  {Alt}}, \bibinfo {author} {\bibfnamefont {T.}~\bibnamefont {Csorgo}},
  \bibinfo {author} {\bibfnamefont {B.}~\bibnamefont {Lorstad}},\ and\ \bibinfo
  {author} {\bibfnamefont {J.}~\bibnamefont {Schmidt-Sorensen}},\ }\bibfield
  {title} {\bibinfo {title} {{Coulomb corrections to the three-body correlation
  function in high-energy heavy ion reactions}},\ }\href
  {https://doi.org/10.1016/S0370-2693(99)00588-2} {\bibfield  {journal}
  {\bibinfo  {journal} {Phys. Lett. B}\ }\textbf {\bibinfo {volume} {458}},\
  \bibinfo {pages} {407} (\bibinfo {year} {1999})},\ \Eprint
  {https://arxiv.org/abs/hep-ph/9812474} {arXiv:hep-ph/9812474} \BibitemShut
  {NoStop}%
\bibitem [{\citenamefont {Liu}\ \emph {et~al.}(1986)\citenamefont {Liu},
  \citenamefont {Beavis}, \citenamefont {Chu}, \citenamefont {Fung},
  \citenamefont {Keane}, \citenamefont {Vandalen},\ and\ \citenamefont
  {Vient}}]{Liu:1986nb}%
  \BibitemOpen
  \bibfield  {author} {\bibinfo {author} {\bibfnamefont {Y.~M.}\ \bibnamefont
  {Liu}}, \bibinfo {author} {\bibfnamefont {D.}~\bibnamefont {Beavis}},
  \bibinfo {author} {\bibfnamefont {S.~Y.}\ \bibnamefont {Chu}}, \bibinfo
  {author} {\bibfnamefont {S.~Y.}\ \bibnamefont {Fung}}, \bibinfo {author}
  {\bibfnamefont {D.}~\bibnamefont {Keane}}, \bibinfo {author} {\bibfnamefont
  {G.~J.}\ \bibnamefont {Vandalen}},\ and\ \bibinfo {author} {\bibfnamefont
  {M.}~\bibnamefont {Vient}},\ }\bibfield  {title} {\bibinfo {title} {{Three
  Pion Correlations in Relativistic Heavy Ion Collisions}},\ }\href
  {https://doi.org/10.1103/PhysRevC.34.1667} {\bibfield  {journal} {\bibinfo
  {journal} {Phys. Rev. C}\ }\textbf {\bibinfo {volume} {34}},\ \bibinfo
  {pages} {1667} (\bibinfo {year} {1986})}\BibitemShut {NoStop}%
\bibitem [{\citenamefont {Binoth}\ and\ \citenamefont
  {Heinrich}(2000)}]{Binoth:2000ps}%
  \BibitemOpen
  \bibfield  {author} {\bibinfo {author} {\bibfnamefont {T.}~\bibnamefont
  {Binoth}}\ and\ \bibinfo {author} {\bibfnamefont {G.}~\bibnamefont
  {Heinrich}},\ }\bibfield  {title} {\bibinfo {title} {{An automatized
  algorithm to compute infrared divergent multiloop integrals}},\ }\href
  {https://doi.org/10.1016/S0550-3213(00)00429-6} {\bibfield  {journal}
  {\bibinfo  {journal} {Nucl. Phys. B}\ }\textbf {\bibinfo {volume} {585}},\
  \bibinfo {pages} {741} (\bibinfo {year} {2000})},\ \Eprint
  {https://arxiv.org/abs/hep-ph/0004013} {arXiv:hep-ph/0004013} \BibitemShut
  {NoStop}%
\bibitem [{\citenamefont {Borowka}\ \emph {et~al.}(2018)\citenamefont
  {Borowka}, \citenamefont {Heinrich}, \citenamefont {Jahn}, \citenamefont
  {Jones}, \citenamefont {Kerner}, \citenamefont {Schlenk},\ and\ \citenamefont
  {Zirke}}]{Borowka:2017idc}%
  \BibitemOpen
  \bibfield  {author} {\bibinfo {author} {\bibfnamefont {S.}~\bibnamefont
  {Borowka}}, \bibinfo {author} {\bibfnamefont {G.}~\bibnamefont {Heinrich}},
  \bibinfo {author} {\bibfnamefont {S.}~\bibnamefont {Jahn}}, \bibinfo {author}
  {\bibfnamefont {S.~P.}\ \bibnamefont {Jones}}, \bibinfo {author}
  {\bibfnamefont {M.}~\bibnamefont {Kerner}}, \bibinfo {author} {\bibfnamefont
  {J.}~\bibnamefont {Schlenk}},\ and\ \bibinfo {author} {\bibfnamefont
  {T.}~\bibnamefont {Zirke}},\ }\bibfield  {title} {\bibinfo {title}
  {{pySecDec: a toolbox for the numerical evaluation of multi-scale
  integrals}},\ }\href {https://doi.org/10.1016/j.cpc.2017.09.015} {\bibfield
  {journal} {\bibinfo  {journal} {Comput. Phys. Commun.}\ }\textbf {\bibinfo
  {volume} {222}},\ \bibinfo {pages} {313} (\bibinfo {year} {2018})},\ \Eprint
  {https://arxiv.org/abs/1703.09692} {arXiv:1703.09692 [hep-ph]} \BibitemShut
  {NoStop}%
\bibitem [{\citenamefont {Borowka}\ \emph {et~al.}(2019)\citenamefont
  {Borowka}, \citenamefont {Heinrich}, \citenamefont {Jahn}, \citenamefont
  {Jones}, \citenamefont {Kerner},\ and\ \citenamefont
  {Schlenk}}]{Borowka:2018goh}%
  \BibitemOpen
  \bibfield  {author} {\bibinfo {author} {\bibfnamefont {S.}~\bibnamefont
  {Borowka}}, \bibinfo {author} {\bibfnamefont {G.}~\bibnamefont {Heinrich}},
  \bibinfo {author} {\bibfnamefont {S.}~\bibnamefont {Jahn}}, \bibinfo {author}
  {\bibfnamefont {S.~P.}\ \bibnamefont {Jones}}, \bibinfo {author}
  {\bibfnamefont {M.}~\bibnamefont {Kerner}},\ and\ \bibinfo {author}
  {\bibfnamefont {J.}~\bibnamefont {Schlenk}},\ }\bibfield  {title} {\bibinfo
  {title} {{A GPU compatible quasi-Monte Carlo integrator interfaced to
  pySecDec}},\ }\href {https://doi.org/10.1016/j.cpc.2019.02.015} {\bibfield
  {journal} {\bibinfo  {journal} {Comput. Phys. Commun.}\ }\textbf {\bibinfo
  {volume} {240}},\ \bibinfo {pages} {120} (\bibinfo {year} {2019})},\ \Eprint
  {https://arxiv.org/abs/1811.11720} {arXiv:1811.11720 [physics.comp-ph]}
  \BibitemShut {NoStop}%
\bibitem [{\citenamefont {Kopylov}\ and\ \citenamefont
  {Podgoretsky}(1975)}]{Kopylov:1975rp}%
  \BibitemOpen
  \bibfield  {author} {\bibinfo {author} {\bibfnamefont {G.~I.}\ \bibnamefont
  {Kopylov}}\ and\ \bibinfo {author} {\bibfnamefont {M.~I.}\ \bibnamefont
  {Podgoretsky}},\ }\bibfield  {title} {\bibinfo {title} {{The Interference of
  Two-Particle States in Particle Physics and Astronomy}},\ }\href@noop {}
  {\bibfield  {journal} {\bibinfo  {journal} {Zh. Eksp. Teor. Fiz.}\ }\textbf
  {\bibinfo {volume} {69}},\ \bibinfo {pages} {414} (\bibinfo {year}
  {1975})}\BibitemShut {NoStop}%
\bibitem [{\citenamefont {Gyulassy}\ \emph {et~al.}(1979)\citenamefont
  {Gyulassy}, \citenamefont {Kauffmann},\ and\ \citenamefont
  {Wilson}}]{Gyulassy:1979yi}%
  \BibitemOpen
  \bibfield  {author} {\bibinfo {author} {\bibfnamefont {M.}~\bibnamefont
  {Gyulassy}}, \bibinfo {author} {\bibfnamefont {S.~K.}\ \bibnamefont
  {Kauffmann}},\ and\ \bibinfo {author} {\bibfnamefont {L.~W.}\ \bibnamefont
  {Wilson}},\ }\bibfield  {title} {\bibinfo {title} {{Pion Interferometry of
  Nuclear Collisions. 1. Theory}},\ }\href
  {https://doi.org/10.1103/PhysRevC.20.2267} {\bibfield  {journal} {\bibinfo
  {journal} {Phys. Rev. C}\ }\textbf {\bibinfo {volume} {20}},\ \bibinfo
  {pages} {2267} (\bibinfo {year} {1979})}\BibitemShut {NoStop}%
\bibitem [{\citenamefont {Andreev}\ \emph {et~al.}(1993)\citenamefont
  {Andreev}, \citenamefont {Plumer},\ and\ \citenamefont
  {Weiner}}]{Andreev:1992pu}%
  \BibitemOpen
  \bibfield  {author} {\bibinfo {author} {\bibfnamefont {I.~V.}\ \bibnamefont
  {Andreev}}, \bibinfo {author} {\bibfnamefont {M.}~\bibnamefont {Plumer}},\
  and\ \bibinfo {author} {\bibfnamefont {R.~M.}\ \bibnamefont {Weiner}},\
  }\bibfield  {title} {\bibinfo {title} {{Quantum statistical approach to
  Bose-Einstein correlations and its experimental implications}},\ }\href
  {https://doi.org/10.1142/S0217751X93001843} {\bibfield  {journal} {\bibinfo
  {journal} {Int. J. Mod. Phys. A}\ }\textbf {\bibinfo {volume} {8}},\ \bibinfo
  {pages} {4577} (\bibinfo {year} {1993})}\BibitemShut {NoStop}%
\bibitem [{\citenamefont {Adam}\ \emph {et~al.}(2016)\citenamefont {Adam} \emph
  {et~al.}}]{ALICE:2015ryj}%
  \BibitemOpen
  \bibfield  {author} {\bibinfo {author} {\bibfnamefont {J.}~\bibnamefont
  {Adam}} \emph {et~al.} (\bibinfo {collaboration} {ALICE}),\ }\bibfield
  {title} {\bibinfo {title} {{Multipion Bose-Einstein correlations in $pp$,
  $p$-Pb, and Pb-Pb collisions at energies available at the CERN Large Hadron
  Collider}},\ }\href {https://doi.org/10.1103/PhysRevC.93.054908} {\bibfield
  {journal} {\bibinfo  {journal} {Phys. Rev. C}\ }\textbf {\bibinfo {volume}
  {93}},\ \bibinfo {pages} {054908} (\bibinfo {year} {2016})},\ \Eprint
  {https://arxiv.org/abs/1512.08902} {arXiv:1512.08902 [nucl-ex]} \BibitemShut
  {NoStop}%
\bibitem [{\citenamefont {Abelev}\ \emph
  {et~al.}(2014{\natexlab{a}})\citenamefont {Abelev} \emph
  {et~al.}}]{ALICE:2013uhj}%
  \BibitemOpen
  \bibfield  {author} {\bibinfo {author} {\bibfnamefont {B.~B.}\ \bibnamefont
  {Abelev}} \emph {et~al.} (\bibinfo {collaboration} {ALICE}),\ }\bibfield
  {title} {\bibinfo {title} {{Two- and three-pion quantum statistics
  correlations in Pb-Pb collisions at $\sqrt{{s}_{NN}} =$ 2.76 TeV at the CERN
  Large Hadron Collider}},\ }\href {https://doi.org/10.1103/PhysRevC.89.024911}
  {\bibfield  {journal} {\bibinfo  {journal} {Phys. Rev. C}\ }\textbf {\bibinfo
  {volume} {89}},\ \bibinfo {pages} {024911} (\bibinfo {year}
  {2014}{\natexlab{a}})},\ \Eprint {https://arxiv.org/abs/1310.7808}
  {arXiv:1310.7808 [nucl-ex]} \BibitemShut {NoStop}%
\bibitem [{\citenamefont {Abelev}\ \emph
  {et~al.}(2014{\natexlab{b}})\citenamefont {Abelev} \emph
  {et~al.}}]{ALICE:2014xrc}%
  \BibitemOpen
  \bibfield  {author} {\bibinfo {author} {\bibfnamefont {B.~B.}\ \bibnamefont
  {Abelev}} \emph {et~al.} (\bibinfo {collaboration} {ALICE}),\ }\bibfield
  {title} {\bibinfo {title} {{Freeze-out radii extracted from three-pion
  cumulants in pp, p\textendash{}Pb and Pb\textendash{}Pb collisions at the
  LHC}},\ }\href {https://doi.org/10.1016/j.physletb.2014.10.034} {\bibfield
  {journal} {\bibinfo  {journal} {Phys. Lett. B}\ }\textbf {\bibinfo {volume}
  {739}},\ \bibinfo {pages} {139} (\bibinfo {year} {2014}{\natexlab{b}})},\
  \Eprint {https://arxiv.org/abs/1404.1194} {arXiv:1404.1194 [nucl-ex]}
  \BibitemShut {NoStop}%
\bibitem [{\citenamefont {Schwartz}(2014)}]{Schwartz:2014sze}%
  \BibitemOpen
  \bibfield  {author} {\bibinfo {author} {\bibfnamefont {M.~D.}\ \bibnamefont
  {Schwartz}},\ }\href@noop {} {\emph {\bibinfo {title} {{Quantum Field Theory
  and the Standard Model}}}}\ (\bibinfo  {publisher} {Cambridge University
  Press},\ \bibinfo {year} {2014})\BibitemShut {NoStop}%
\bibitem [{\citenamefont {Heinrich}(2008)}]{Heinrich:2008si}%
  \BibitemOpen
  \bibfield  {author} {\bibinfo {author} {\bibfnamefont {G.}~\bibnamefont
  {Heinrich}},\ }\bibfield  {title} {\bibinfo {title} {{Sector
  Decomposition}},\ }\href {https://doi.org/10.1142/S0217751X08040263}
  {\bibfield  {journal} {\bibinfo  {journal} {Int. J. Mod. Phys. A}\ }\textbf
  {\bibinfo {volume} {23}},\ \bibinfo {pages} {1457} (\bibinfo {year}
  {2008})},\ \Eprint {https://arxiv.org/abs/0803.4177} {arXiv:0803.4177
  [hep-ph]} \BibitemShut {NoStop}%
\bibitem [{\citenamefont {Heinrich}\ \emph {et~al.}(2022)\citenamefont
  {Heinrich}, \citenamefont {Jahn}, \citenamefont {Jones}, \citenamefont
  {Kerner}, \citenamefont {Langer}, \citenamefont {Magerya}, \citenamefont
  {P\"oldaru}, \citenamefont {Schlenk},\ and\ \citenamefont
  {Villa}}]{Heinrich:2021dbf}%
  \BibitemOpen
  \bibfield  {author} {\bibinfo {author} {\bibfnamefont {G.}~\bibnamefont
  {Heinrich}}, \bibinfo {author} {\bibfnamefont {S.}~\bibnamefont {Jahn}},
  \bibinfo {author} {\bibfnamefont {S.~P.}\ \bibnamefont {Jones}}, \bibinfo
  {author} {\bibfnamefont {M.}~\bibnamefont {Kerner}}, \bibinfo {author}
  {\bibfnamefont {F.}~\bibnamefont {Langer}}, \bibinfo {author} {\bibfnamefont
  {V.}~\bibnamefont {Magerya}}, \bibinfo {author} {\bibfnamefont
  {A.}~\bibnamefont {P\"oldaru}}, \bibinfo {author} {\bibfnamefont
  {J.}~\bibnamefont {Schlenk}},\ and\ \bibinfo {author} {\bibfnamefont
  {E.}~\bibnamefont {Villa}},\ }\bibfield  {title} {\bibinfo {title}
  {{Expansion by regions with pySecDec}},\ }\href
  {https://doi.org/10.1016/j.cpc.2021.108267} {\bibfield  {journal} {\bibinfo
  {journal} {Comput. Phys. Commun.}\ }\textbf {\bibinfo {volume} {273}},\
  \bibinfo {pages} {108267} (\bibinfo {year} {2022})},\ \Eprint
  {https://arxiv.org/abs/2108.10807} {arXiv:2108.10807 [hep-ph]} \BibitemShut
  {NoStop}%
\bibitem [{\citenamefont {Bruns}\ and\ \citenamefont
  {Ichim}(2010)}]{BRUNS20101098}%
  \BibitemOpen
  \bibfield  {author} {\bibinfo {author} {\bibfnamefont {W.}~\bibnamefont
  {Bruns}}\ and\ \bibinfo {author} {\bibfnamefont {B.}~\bibnamefont {Ichim}},\
  }\bibfield  {title} {\bibinfo {title} {Normaliz: Algorithms for affine
  monoids and rational cones},\ }\href
  {https://doi.org/https://doi.org/10.1016/j.jalgebra.2010.01.031} {\bibfield
  {journal} {\bibinfo  {journal} {Journal of Algebra}\ }\textbf {\bibinfo
  {volume} {324}},\ \bibinfo {pages} {1098} (\bibinfo {year} {2010})},\
  \bibinfo {note} {computational Algebra}\BibitemShut {NoStop}%
\bibitem [{\citenamefont {Ellis}(2017)}]{Ellis:2016jkw}%
  \BibitemOpen
  \bibfield  {author} {\bibinfo {author} {\bibfnamefont {J.}~\bibnamefont
  {Ellis}},\ }\bibfield  {title} {\bibinfo {title} {{TikZ-Feynman: Feynman
  diagrams with TikZ}},\ }\href {https://doi.org/10.1016/j.cpc.2016.08.019}
  {\bibfield  {journal} {\bibinfo  {journal} {Comput. Phys. Commun.}\ }\textbf
  {\bibinfo {volume} {210}},\ \bibinfo {pages} {103} (\bibinfo {year}
  {2017})},\ \Eprint {https://arxiv.org/abs/1601.05437} {arXiv:1601.05437
  [hep-ph]} \BibitemShut {NoStop}%
\bibitem [{\citenamefont {Akkelin}\ \emph {et~al.}(2002)\citenamefont
  {Akkelin}, \citenamefont {Lednicky},\ and\ \citenamefont
  {Sinyukov}}]{Akkelin:2001nd}%
  \BibitemOpen
  \bibfield  {author} {\bibinfo {author} {\bibfnamefont {S.~V.}\ \bibnamefont
  {Akkelin}}, \bibinfo {author} {\bibfnamefont {R.}~\bibnamefont {Lednicky}},\
  and\ \bibinfo {author} {\bibfnamefont {Y.~M.}\ \bibnamefont {Sinyukov}},\
  }\bibfield  {title} {\bibinfo {title} {{Correlation search for coherent pion
  emission in heavy ion collisions}},\ }\href
  {https://doi.org/10.1103/PhysRevC.65.064904} {\bibfield  {journal} {\bibinfo
  {journal} {Phys. Rev. C}\ }\textbf {\bibinfo {volume} {65}},\ \bibinfo
  {pages} {064904} (\bibinfo {year} {2002})},\ \Eprint
  {https://arxiv.org/abs/nucl-th/0107015} {arXiv:nucl-th/0107015} \BibitemShut
  {NoStop}%
\bibitem [{\citenamefont {Smirnov}\ and\ \citenamefont
  {Chuharev}(2020)}]{Smirnov:2019qkx}%
  \BibitemOpen
  \bibfield  {author} {\bibinfo {author} {\bibfnamefont {A.~V.}\ \bibnamefont
  {Smirnov}}\ and\ \bibinfo {author} {\bibfnamefont {F.~S.}\ \bibnamefont
  {Chuharev}},\ }\bibfield  {title} {\bibinfo {title} {{FIRE6: Feynman Integral
  REduction with Modular Arithmetic}},\ }\href
  {https://doi.org/10.1016/j.cpc.2019.106877} {\bibfield  {journal} {\bibinfo
  {journal} {Comput. Phys. Commun.}\ }\textbf {\bibinfo {volume} {247}},\
  \bibinfo {pages} {106877} (\bibinfo {year} {2020})},\ \Eprint
  {https://arxiv.org/abs/1901.07808} {arXiv:1901.07808 [hep-ph]} \BibitemShut
  {NoStop}%
\bibitem [{\citenamefont {von Manteuffel}\ \emph {et~al.}(2015)\citenamefont
  {von Manteuffel}, \citenamefont {Panzer},\ and\ \citenamefont
  {Schabinger}}]{vonManteuffel:2014qoa}%
  \BibitemOpen
  \bibfield  {author} {\bibinfo {author} {\bibfnamefont {A.}~\bibnamefont {von
  Manteuffel}}, \bibinfo {author} {\bibfnamefont {E.}~\bibnamefont {Panzer}},\
  and\ \bibinfo {author} {\bibfnamefont {R.~M.}\ \bibnamefont {Schabinger}},\
  }\bibfield  {title} {\bibinfo {title} {{A quasi-finite basis for multi-loop
  Feynman integrals}},\ }\href {https://doi.org/10.1007/JHEP02(2015)120}
  {\bibfield  {journal} {\bibinfo  {journal} {JHEP}\ }\textbf {\bibinfo
  {volume} {02}},\ \bibinfo {pages} {120}},\ \Eprint
  {https://arxiv.org/abs/1411.7392} {arXiv:1411.7392 [hep-ph]} \BibitemShut
  {NoStop}%
\bibitem [{\citenamefont {Smirnov}(2012)}]{Smirnov:2012gma}%
  \BibitemOpen
  \bibfield  {author} {\bibinfo {author} {\bibfnamefont {V.~A.}\ \bibnamefont
  {Smirnov}},\ }\href {https://doi.org/10.1007/978-3-642-34886-0} {\emph
  {\bibinfo {title} {{Analytic tools for Feynman integrals}}}},\ Vol.\ \bibinfo
  {volume} {250}\ (\bibinfo {year} {2012})\BibitemShut {NoStop}%
\end{thebibliography}%

\end{document}